\documentclass[conference, hidelinks]{IEEEtran}

\usepackage{amsmath,amssymb,amsfonts}
\usepackage{array}
\usepackage{textcomp}
\usepackage{url}
\usepackage{verbatim}
\usepackage{graphicx}
\usepackage{cite}
\usepackage{xcolor}
\usepackage{orcidlink}    
\usepackage[acronym,shortcuts]{glossaries}
\usepackage{bm}
\usepackage[font=footnotesize]{caption}
\usepackage{subcaption}
\usepackage{relsize}
\usepackage{soul}
\usepackage{algorithm, algpseudocode}
\usepackage{algcompatible}
\usepackage{lipsum}
\usepackage{mathtools}
\usepackage{booktabs}
\usepackage[bottom]{footmisc} 
\usepackage{tabularx}
\usepackage{multirow}
\usepackage{stfloats}
\usepackage{amsthm}
\def\BibTeX{{\rm B\kern-.05em{\sc i\kern-.025em b}\kern-.08em
T\kern-.1667em\lower.7ex\hbox{E}\kern-.125emX}}

\newcommand{\trans}[0]{^{\mathsf{T}}}

\newcommand{\herm}[0]{^{\mathsf{H}}}

\newacronym{OFDM}{OFDM}{orthogonal frequency division multiplexing}
\newacronym{DFT-s-OFDM}{DFT-s-OFDM}{DFT-spread OFDM}
\newacronym{FBMC}{FBMC}{filter bank multi-carrier}
\newacronym{GFDM}{GFDM}{generalized frequency division multiplexing}
\newacronym{OCDM}{OCDM}{orthogonal chirp division multiplexing}
\newacronym{AFDM}{AFDM}{affine frequency division multiplexing}
\newacronym{OTFS}{OTFS}{orthogonal time frequency space}
\newacronym{T-OTFS}{T-OTFS}{transcendentally-rotated OTFS}
\newacronym{ODDM}{ODDM}{orthogonal delay-Doppler division multiplexing}
\newacronym{SWHM}{SWHM}{sparse Walsh-Hadamard multiplexing}
\newacronym{FMCW}{FMCW}{frequency modulated continuous wave}
\newacronym{OTSM}{OTSM}{orthogonal time sequency modulation}
\newacronym{MMSE}{MMSE}{miminum mean squared error}
\newacronym{ICI}{ICI}{inter-carrier interference}
\newacronym{RMSE}{RMSE}{root-mean-squared error}
\newacronym{CPP}{CPP}{chirp-periodic prefix}
\newacronym{PDA}{PDA}{probabilistic data association}
\newacronym{LCT}{LCT}{linear canonical transform}
\newacronym{ZT}{ZT}{Zak transform}
\newacronym{DZT}{DZT}{discrete Zak transform}
\newacronym{IDZT}{IDZT}{inverse discrete Zak transform}
\newacronym{CP-DAFT}{CP-DAFT}{chirp-permuted discrete affine Fourier transform}
\newacronym{CP-IDAFT}{CP-IDAFT}{chirp-permuted inverse discrete affine Fourier transform}
\newacronym{FT}{FT}{Fourier transform}
\newacronym{IFT}{IFT}{inverse Fourier transform}
\newacronym{DFT}{DFT}{discrete Fourier transform}
\newacronym{IDFT}{IDFT}{inverse discrete Fourier transform}
\newacronym{AFT}{AFT}{affine Fourier transform}
\newacronym{DAFT}{DAFT}{discrete affine Fourier transform}
\newacronym{IDAFT}{IDAFT}{inverse discrete affine Fourier transform}
\newacronym{SFFT}{SFFT}{symplectic finite Fourier transform}
\newacronym{ITN}{ITN}{intelligent traffic network}
\newacronym{ISFFT}{ISFFT}{inverse symplectic finite Fourier transform}
\newacronym{HT}{HT}{Heisenberg transform}
\newacronym{SISO}{SISO}{single-input single-output}
\newacronym{MIMO}{MIMO}{multiple-input multiple-output}
\newacronym{WT}{WT}{Wigner transform}
\newacronym{frFT}{frFT}{fractional Fourier transform}
\newacronym{IfrFT}{IfrFT}{inverse fractional Fourier transform}
\newacronym{fnT}{fnT}{Fresnel transform}
\newacronym{IfnT}{IfnT}{inverse Fresnel transform}
\newacronym{LT}{LT}{Laplace transform}
\newacronym{ILT}{ILT}{inverse Laplace transform}
\newacronym{ISAC}{ISAC}{integrated sensing and communications}
\newacronym{JCAS}{JCAS}{joint communications and sensing}
\newacronym{EM}{EM}{electromagnetic}
\newacronym{CP}{CP}{chirp-permuted}
\newacronym{B5G}{B5G}{beyond fifth generation}
\newacronym{3GPP}{3GPP}{$3^{\rm{rd}}$ generation partnership project}           
\newacronym{4G}{4G}{fourth generation}                                
\newacronym{5G}{5G}{fifth generation}                               
\newacronym{6G}{6G}{sixth generation}
\newacronym{CPIM}{CPIM}{chirp-permutation index modulation}
\newacronym{LTV}{LTV}{linear time-variant}
\newacronym{LTI}{LTI}{linear time-invariant}
\newacronym{LTVM}{LTVM}{linear time-variant multipath}
\newacronym{TV}{TV}{time-variant}
\newacronym{TI}{TI}{time-invariant}
\newacronym{1D}{1D}{one-dimensional}
\newacronym{2D}{2D}{two-dimensional}
\newacronym{3D}{3D}{three-dimensional}
\newacronym{NTN}{NTN}{non-terrestrial network}
\newacronym{LEO}{LEO}{low earth orbit}
\newacronym{IoT}{IoT}{Internet-of-Things}
\newacronym{mmWave}{mmWave}{millimeter-wave}
\newacronym{THz}{THz}{Terahertz}
\newacronym{V2X}{V2X}{vehicle-to-everything}
\newacronym{RCC}{RCC}{radar-communication coexistence}
\newacronym{C-V2X}{C-V2X}{Cellular-V2X}                             
\newacronym{NR}{NR}{New Radio}                  
\newacronym{Tbps}{Tbps}{Terabits-per-second}
\newacronym{ms}{ms}{millisecond}
\newacronym{UAV}{UAV}{unmanned aerial vehicle}
\newacronym{CFO}{CFO}{carrier frequency offset}
\newacronym{ITS}{ITS}{intelligent transportation system}
\newacronym{SAGIN}{SAGIN}{space-air-ground integrated network}
\newacronym{DTN}{DTN}{digital twin network}
\newacronym{ETSI}{ETSI}{European Telecommunications Standards Institute}
\newacronym{EHF}{EHF}{extremely high-frequency}
\newacronym{BER}{BER}{bit-error-rate}
\newacronym{SotA}{SotA}{state-of-the-art}
\newacronym{DoF}{DoF}{degrees-of-freedom}
\newacronym{UE}{UE}{user equipment}
\newacronym{AP}{AP}{access point}
\newacronym{CIR}{CIR}{channel impulse response}
\newacronym{CSI}{CSI}{channel state information}
\newacronym{SNR}{SNR}{signal-to-noise ratio}
\newacronym{AWGN}{AWGN}{additive white Gaussian noise}
\newacronym{ML}{ML}{maximum likelihood}
\newacronym{OOB}{OOB}{out-of-band}
\newacronym{MX}{MX}{multiplexing}
\newacronym{DMX}{DMX}{demultiplexing}
\newacronym{AoA}{AoA}{angle-of-arrival}
\newacronym{AoD}{AoD}{angle-of-departure}
\newacronym{5GAA}{5GAA}{5G Automotive Association}                               
\newacronym{6G-IA}{6G-IA}{6G Smart Networks and Services Industry Association}
\newacronym{MUSIC}{MUSIC}{MUltiple SIgnal Classification}
\newacronym{ESPRIT}{ESPRIT}{estimation of signal parameters via rotational invariance techniques}
\newacronym{DP}{DP}{detection problem}
\newacronym{EP}{EP}{estimation problem}
\newacronym{KPI}{KPI}{key performance indicator}
\newacronym{ISI}{ISI}{inter-symbol interference}
\newacronym{TVIRF}{TVIRF}{time-variant impulse response function}
\newacronym{TVTF}{TVTF}{time-variant transfer function}
\newacronym{DVIRF}{DVIRF}{Doppler-variant impulse response function}
\newacronym{DVTF}{DVTF}{Doppler-variant transfer function}
\newacronym{IM}{IM}{index modulation}
\newacronym{DDSF}{DDSF}{delay-Doppler spread function}
\newacronym{PAPR}{PAPR}{peak-to-average power ratio}
\newacronym{RCS}{RCS}{radar cross-section}
\newacronym{PIR}{PIR}{peak interference residual}
\newacronym{PSR}{PSR}{peak-to-sidelobe ratio}
\newacronym{PA}{PA}{power amplifier}
\newacronym{RF}{RF}{radio frequency}
\newacronym{JCDE}{JCDE}{joint channel and data estimation}
\newacronym{AF}{AF}{ambiguity function}

\newacronym{i.i.d.}{i.i.d.}{independent and identically distributed}
\newacronym{PSLR}{PSLR}{peak-to-sidelobe ratio}
\newacronym{ISLR}{ISLR}{integrated sidelobe ratio}
\newacronym{CDF}{CDF}{cumulative distribution function}
\newacronym{CCDF}{CCDF}{complementary cumulative distribution function}
\newacronym{CP-AFDM}{CP-AFDM}{chirp-permuted AFDM}

\begin{document}

\title{Normalized Ambiguity Function Characteristics of OFDM, OTFS, AFDM, and CP-AFDM for ISAC}

\author{\IEEEauthorblockN{Hyeon Seok Rou\textsuperscript{\orcidlink{0000-0003-3483-7629}} and Giuseppe Thadeu Freitas de Abreu\textsuperscript{\orcidlink{0000-0002-5018-8174}}}
\IEEEauthorblockA{
\textit{Constructor University, Bremen, Germany}}
\IEEEauthorblockA{{[hrou, gabreu]@constructor.university}}
\IEEEauthorblockA{\\[-2ex]}
}

\maketitle

\begin{abstract}
This paper presents a unified and system-agnostic analysis of the \ac{AF} characteristics of four representative multicarrier waveforms, \ac{OFDM}, \ac{OTFS}, \ac{AFDM}, and \ac{CP-AFDM}, which are considered as key candidates for enabling \ac{ISAC} in future \ac{6G} networks.
The \ac{AF} of each waveform is obtained directly from its discrete-time definition and enhanced via ideal fractional interpolation, enabling precise characterization of its continuous-time delay-Doppler response.
Two signaling modes are examined: a communication-oriented case with random information symbols suitable only for monostatic scenarios, and a sensing-oriented case with fixed unimodular symbols suitable for general multi-static scenarios.
Furthermore, the \acp{AF} and the ambiguity metrics including the 3dB mainlobe width, \ac{PSLR}, and \ac{ISLR}, are evaluated in normalized delay-Doppler units, enabling direct translation to any physical system configuration defined by bandwidth, sampling frequency, or symbol duration, while ensuring straightforward and consistent comparison across waveforms.
The results establish a consistent benchmark for comparing waveform sensing capabilities in \ac{ISAC} design, consolidating known behaviors: \ac{OFDM} exhibits excellent delay resolution and sidelobe behavior but poor Doppler response, whereas advanced waveforms achieve improved balance between delay and Doppler resolution with varying sidelobe characteristics.
The simulation code of the smooth \acp{AF}, is openly shared to promote reproducibility and support future \ac{ISAC} waveform research.
\end{abstract}


\glsresetall

\section{Introduction}
\label{sec:introduction}

Future wireless networks are envisioned to unify communication and sensing functionalities within a single radio framework, enabling the simultaneous transmission of information and perception of the surrounding environment using a common waveform, hardware platform, and signal processing chain~\cite{Chowdhury_6G,Wang_6G}. 
This emerging paradigm, popularly known as \ac{ISAC}, is one of the foundational pillars in the outlined \ac{6G} system definitions and requirements~\cite{10529727,9737357}.

Radar-centric \ac{ISAC} approaches, which employ waveforms originally optimized for radar operation, such as \ac{FMCW}, have naturally been investigated for this by embedding information into such waveforms while preserving the sensing capacity~\cite{10149620,ma2021spatial}.
However, they are constrained by their limited compatibility with existing communication standards, and often fail to meet the high data rate requirements of \ac{6G} \cite{zhang2021overview}.    

Given these limitations, communication-centric \ac{ISAC} approaches, have emerged as natural candidates for seamless integration, owing to their inherently high spectral efficiency and communications performance, compatibility with existing standards, and flexibility in waveform design~\cite{koivunen2024multicarrier,10769778}. 

Among such waveforms, \ac{OFDM} serves as a natural baseline, being the dominant modulation scheme in current wireless systems, in which many \ac{ISAC} studies have been conducted~\cite{braun2014ofdm,dai2025tutorial}, highlighting its excellent delay resolution (target range) and sidelobe behavior \cite{liu2025cp}, but severe limitations in Doppler resolution (target velocity).
Alternatively, several advanced waveforms proposed for \ac{6G}, such as \ac{OTFS} \cite{hadani2017orthogonal, wei2021orthogonal}, \ac{AFDM} \cite{10087310,rou2025affine}, and \ac{CP-AFDM} \cite{rou2025CPAFDM} exhibit inherently strong delay-Doppler localization, thereby implying highly attractive properties for sensing applications, particularly in the velocity estimation.

This motivates a unified and systematic comparison of waveform characteristics and sensing performance, for which we leverage the \ac{AF}: an intrinsic property of the transmit waveform independent of the propagation channel and system configuration, which directly reflects its range-velocity resolution and sidelobe behavior through metrics such as the mainlobe width, \ac{PSLR}, and \ac{ISLR} \cite{richards2010principles, li2008mimo}.

While the \ac{AF} of individual waveforms has been investigated in many prior studies~\cite{yin2025ambiguity,wang2025optimal,chong2025delay}, these analyses have largely been conducted in isolation and lack a common framework for consistent comparison across both the delay and Doppler domains in high definition analogous to the true continuous-time behavior.
Therefore, we harmonize the analysis across multiple candidate waveforms by establishing a physical system-agnostic discrete-time \ac{AF} evaluation framework that employs normalized delay and Doppler coordinates, enabling simple translation to any practical system configuration, given its physical bandwidth, sampling frequency, or symbol duration. 

Furthermore, to approximate the true continuous behavior, the \ac{AF} is refined through optimal interpolation in delay and oversampling in Doppler, yielding smooth, high-resolution representations suitable for both visualization and quantitative metric assessment. 
Finally, two signaling modes are considered for comparison across all waveforms: a general \ac{ISAC} mode with random information symbols, and a sensing-focused mode using deterministic unimodular symbols.

\section{System Model}
\label{sec:system_model}

Consider a block of $N$ complex information symbols $\mathbf{x} \triangleq [x_0, x_1, \ldots, x_{N-1}]^\mathsf{T} \in \mathcal{X}^{N\times 1}$ drawn from a normalized $M$-ary constellation $\mathcal{X} \subset \mathbb{C}$ with cardinality $M$, i.e., $M$-QAM constellation. 
The encoding of the symbol vector $\mathbf{x}$ into a discrete-time transmit signal $\mathbf{s} \triangleq [s_0, s_1, \ldots, s_{N-1}]^\mathsf{T} \in \mathbb{C}^{N\times 1}$ is performed by a unitary linear modulator $\mathbf{M} \in \mathbb{C}^{N \times N}$.
In other words, the discrete-time transmit signal vector $\mathbf{s} \in \mathbb{C}^{N \times 1}$ is given by
\begin{equation}
\mathbf{s} \triangleq \mathbf{M} \mathbf{x} \in \mathbb{C}^{N \times 1}.
\end{equation}

In light of the above, we consider four representative waveforms, namely \ac{OFDM}, \ac{OTFS}, \ac{AFDM}, and \ac{CP-AFDM}, whose respective modulation matrices $\mathbf{M}^{\text{OFDM}}$, $\mathbf{M}^{\text{OTFS}}$, $\mathbf{M}^{\text{AFDM}}$, and $\mathbf{M}^{\text{CP-AFDM}}$ are based on the normalized $N$-point \ac{DFT} matrix $\mathbf{F}_N \in \mathbb{C}^{N\times N}$, whose $(i,j)$-th entry is given by $[\mathbf{F}_N]_{i,n} = \tfrac{1}{\sqrt{N}} e^{-j 2\pi \frac{in}{N}}$. 
The resulting baseband transmit signals and the respective modulator structure for the four waveforms are summarized below.

\subsection{Orthogonal Frequency-Division Multiplexing (OFDM)}

For conventional \ac{OFDM}, the data symbols are mapped directly onto orthogonal subcarriers through the inverse discrete Fourier transform (IDFT), i.e., $\mathbf{M}^\mathrm{OFDM} = \mathbf{F}_N\herm$, yielding the transmit signal
\begin{equation}
\mathbf{s}^{\text{OFDM}}
= \mathbf{F}_N\herm \mathbf{x} \in \mathbb{C}^{N \times 1}.
\label{eq:ofdm_tx}
\end{equation}

\subsection{Orthogonal Time-Frequency Space Modulation (OTFS)}

The \ac{OTFS} waveform is designed to provide robustness against doubly-dispersive channels by utilizing the delay-Doppler domain for information symbol placement, and have been shown to provide significant performance gains in high-mobility scenarios~\cite{hadani2017orthogonal,wei2021orthogonal}.
Therefore, the information symbols are directly placed on the two-dimensional delay-Doppler grid $\mathbf{X} \in \mathcal{X}^{K \times L}$, with $N = LK$, which undergoes a two-dimensional \ac{ISFFT} and a \ac{HT} to yield the discrete-time transmit signal.
Assuming rectangular pulse shaping, the discrete-time \ac{OTFS} transmit signal can be expressed as
\begin{equation}
\mathbf{s}^{\text{OTFS}} = \mathrm{vec} \Big( \mathbf{F}_K\herm \big(\mathbf{F}_K \mathbf{X} \mathbf{F}_L\herm\big) \Big) = (\mathbf{F}_L\herm \otimes \mathbf{I}_K^{\vphantom{x}}) \mathbf{x} \in \mathbb{C}^{N \times 1},
\end{equation}
where $\mathrm{vec}(\cdot)$ denotes the vectorization operator and $\otimes$ is the Kronecker product, and therefore $\mathbf{M}^\mathrm{OTFS} = (\mathbf{F}_L\herm \otimes \mathbf{I}_K^{\vphantom{x}})$.

\subsection{Affine Frequency-Division Multiplexing (AFDM)}

On the other hand, the \ac{AFDM} was proposed also to provide robustness against doubly-dispersive channels, by exploiting chirp-based subcarriers, and has been shown to guarantee full diversity in doubly-dispersive channels and therefore excellent performance in high-mobility scenarios~\cite{Bemani_AFDM23,rou2025affine}.

The \ac{AFDM} waveform generalizes \ac{OFDM} by extending the pure-tone subcarriers into a chirp-based subcarrier, spreading the symbols in the time-frequency domain.
The modulation is concisely achieved through the \ac{IDAFT}, with two tunable chirp parameters $c_1, c_2 \in \mathbb{R}$, which control the chirp rates.
Given the above, the discrete-time \ac{AFDM} transmit signal is given by
\begin{equation}
\mathbf{s}^{\text{AFDM}}
= \mathbf{\Lambda}_{c_1}\herm \mathbf{F}_N\herm \mathbf{\Lambda}_{c_2}\herm \mathbf{x} \in \mathbb{C}^{N \times 1},
\label{eq:afdm_tx}
\end{equation}
with $\mathbf{M}^\mathrm{AFDM} =  \mathbf{\Lambda}_{c_1}\herm \mathbf{F}_N\herm \mathbf{\Lambda}_{c_2}\herm$, and $\mathbf{\Lambda}_{c_i} = \mathrm{diag}(\boldsymbol{\lambda}_{c_i})$ are the diagonal chirp matrices with chirp sequences
\begin{equation}
\boldsymbol{\lambda}_{c_i} =
\big[e^{-j2\pi c_i (0)^2}, \ldots, e^{-j2\pi c_i (N-1)^2}\big]\trans \in \mathbb{C}^{N \times 1},
\end{equation}
for both $c_1$ and $c_2$.

\subsection{Chirp-Permuted AFDM (CP-AFDM)}

Recently, the \ac{CP-AFDM} waveform was proposed to extend the \ac{AFDM} by permuting the second chirp sequence $\boldsymbol{\lambda}_{c_2}$ prior to modulation, thereby introducing an additional design degree of freedom to be utilized for additional functionalities, such as \ac{IM} \cite{rou2024afdm} or physical layer security \cite{rou2025chirp}, while retaining the beneficial delay-Doppler characteristics of the conventional \ac{AFDM} \cite{rou2025CPAFDM}.

Let $\mathbf{\Pi}_i \in \mathbb{C}^{N\times N}$ 
be a permutation matrix corresponding to index $i \in \{1,\ldots,N!\}$, where the permutation index $i$ follows the lexicographic order of all possible permutations of length $N$.
Then, by permuting the chirp sequence of the second diagonal chirp matrix as
\begin{equation}
\mathbf{\Lambda}_{c_2,i}
= \mathrm{diag}(\mathbf{\Pi}_i \boldsymbol{\lambda}_{c_2}) \in \mathbb{C}^{N \times N},
\end{equation}
the resulting discrete-time signal of the chirp-permuted \ac{AFDM} with order $i$ is simply given by
\begin{equation}
\mathbf{s}^{\text{CP-AFDM},i}
= \mathbf{\Lambda}_{c_1}\herm \mathbf{F}_N\herm \mathbf{\Lambda}_{c_2,i}\herm \mathbf{x} \in \mathbb{C}^{N \times 1},
\label{eq:cpafdm_tx}
\end{equation}

It should be noted that the modulator matrix $\mathbf{M}^{\text{CP-AFDM},i} = \mathbf{\Lambda}_{c_1}\herm \mathbf{F}_N\herm \mathbf{\Lambda}_{c_2,i}\herm$, \ac{CP-IDAFT}, remains unitary for all permutation order $i \in \{1,\ldots,N\!\}$.

\section{Ambiguity Function Analysis}
\label{sec:ambiguity}

Given the above discrete-time transmit signal models of the four candidate waveforms, we now present a unified framework for directly evaluating their \ac{AF} characteristics in the normalized delay-Doppler domain, enabling direct comparison and translation to arbitrary system parameters.

The \ac{AF} is a fundamental descriptor of a signal's joint delay-Doppler resolution, and hence a key analytical tool for \ac{ISAC} waveform design and evaluation, especially for non-radar-optimized waveforms.
For a continuous-time baseband waveform $s(t)$, it is defined as
\begin{equation}
A(\tau,\nu) 
= \int_{-\infty}^{\infty} s(t)\, s^{*}(t-\tau)\,
e^{-j 2\pi \nu t}\, \mathrm{d}t,
\label{eq:AF_cont}
\end{equation}
which measures the correlation between $s(t)$ and a version of itself shifted in time and frequency. 
The mainlobe of $|A(\tau,\nu)|$ determines the achievable delay and Doppler resolution, while sidelobes quantify potential ambiguities or interference among multiple reflections, directly affecting sensing accuracy and false detection rates.

In this work, the discrete-time \ac{AF} is computed directly from the finite-length transmit vector 
$\mathbf{s}$, as described in the previous section, whose time samples are assumed to have sampling period $T_s = 1/f_s$, with sampling freqeuncy $f_s$.
Then, the discrete \ac{AF} is expressed as
\begin{equation}
A[\ell,k] = \sum_{n=0}^{N-1} s[n]\, s^{*}[n-\ell]\, e^{-j 2\pi f\frac{n}{N}},
\label{eq:AF_discrete}
\end{equation}
where $\ell$ and $f$ denote the integer delay and Doppler indices, respectively, yielding the exact finite-sum realization of~\eqref{eq:AF_cont}. 

Here, normalized variables can be defined as
\begin{equation}
\tau_{\mathrm{norm}} = \frac{\ell}{N}, 
\qquad
\nu_{\mathrm{norm}} = \frac{f}{f_s},
\label{eq:AF_norm_axes}
\end{equation}
so that $A(\tau_{\mathrm{norm}},\nu_{\mathrm{norm}})$ is independent of sampling rate or bandwidth. 
In practice, $\tau_{\mathrm{norm}} \in [-1,1]$ spans the full unambiguous delay range for a signal of length $N$, while $\nu_{\mathrm{norm}} \in [-0.5,0.5]$ corresponds to the Nyquist-limited Doppler interval (cycles per sample), fully representing the centrally meaningful region without aliasing. 
This normalized region fully captures the intrinsic delay-Doppler characteristics of each waveform, independent of specific system parameters, and can be directly mapped to the corresponding physical quantities (time delay in seconds and Doppler shift in hertz) using
\begin{equation}
\tau_{\mathrm{phys}} = \tau_{\mathrm{norm}}\, T = \ell\, T_s,
\qquad
\nu_{\mathrm{phys}} = \nu_{\mathrm{norm}}\, \frac{1}{T_s},
\label{eq:AF_phys_scale}
\end{equation}
where $T = N T_s$ denotes the total symbol duration of the discrete block.  

On top of this, to achieve smooth, high-resolution \ac{AF} surfaces, both delay and Doppler axes are refined through oversampling and interpolation. 
In the time domain, the discrete signal is reconstructed by an ideal sinc interpolation via the sampling theorem
\begin{equation}
s_{\mathrm{interp}}[t]
= \sum_{n=0}^{N-1} s[n]\,
\mathrm{sinc}\!\left( \frac{t-n}{T_s} \right),
\label{eq:sinc_interp_time}
\end{equation}
which enables the evaluation of the \ac{AF} at fractional delay values with resolution $\Delta \tau = T_s / O_{\tau}$, where $O_{\tau}$ is the delay oversampling factor.

For numerical implementation, however, the infinite sinc kernel is truncated and windowed to yield a practical {windowed-sinc} interpolation of finite support $2L_h$ samples:
\begin{equation}
s_{\mathrm{interp}}(t)
\approx
\sum_{n = n_0 - L_h}^{n_0 + L_h}
s[n]\,
w[n - n_0]\,
\mathrm{sinc}\!\left( \frac{t - nT_s}{T_s} \right),
\label{eq:windowed_sinc_interp}
\end{equation}
where $w[\cdot]$ is a smooth tapering window (e.g., Hann or Kaiser) that mitigates truncation-induced sidelobes.  

Similarly, the Doppler dimension is likewise densified by evaluating \eqref{eq:AF_discrete} on a dense frequency grid $\nu \in [-\nu_{\max}, +\nu_{\max}]$ with resolution $\Delta \nu = 1/(O_{\nu} N)$, where $O_{\nu}$ is the Doppler oversampling factor.
This continuous, oversampled representation provides a numerically stable and theoretically accurate approximation of \eqref{eq:AF_cont}, which enables a high-definition comparison of the waveform \acp{AF}, and precise extraction of key ambiguity metrics in the discrete domain.

Furthermore, while the \ac{AF} is trivially in two dimensions, one-dimensional \ac{AF} cross-sections are often more interpretable and relevant for practical sensing applications, as they directly reflect the waveform's resolution and ambiguity characteristics in either delay or Doppler.
Namely, the {zero-Doppler cut} $A_{\tau}[\ell] = A[\ell,0]$, which reflects delay resolution, and the {zero-delay cut} $A_{\nu}[k] = A[0,k]$, 
which reflects Doppler resolution. 
Both are obtained from the energy-normalized \ac{AF} (i.e., normalized to unit peak) after interpolation, and are evaluated within the aforementioned normalized regions ($\tau_{\mathrm{norm}} \in [-1,1]$, $\nu_{\mathrm{norm}} \in [-0.5,0.5]$), ensuring complete coverage of the waveform's unambiguous sensing characteristics.

From the normalized ambiguity function cuts in zero-delay and zero-Doppler, three quantitative measures are derived to characterize the waveform behavior in delay and Doppler, respectively.
This enables a comprehensive assessment of the waveform's sensing capabilities in both range and velocity estimation, which are critical for \ac{ISAC} applications, rather than qualitatively comparing the \acp{AF} of different waveforms.

The first is the \textit{3\,dB mainlobe width}, which defines the half-power region surrounding the AF peak on either axis (delay or Doppler), and is computed as the width between the two points $\xi_{\pm}$ where the \ac{AF} magnitude drops to $\frac{1}{\sqrt{2}}$ of its peak value, i.e.,
\begin{equation}
|A(\xi_{\pm})| = \frac{1}{\sqrt{2}}, 
\qquad 
\Delta \xi_{\mathrm{norm}} = \xi_{+} - \xi_{-},
\label{eq:mainlobe_width}
\end{equation}
where $\xi$ denotes either the normalized delay ($\tau_{\mathrm{norm}}$) or Doppler ($\nu_{\mathrm{norm}}$) variable. 

Note that the width is measured in terms of the normalized delay or Doppler units of eq. \eqref{eq:AF_cont}, such that it can be translated to the practical resolutions given the system parameters using eq. \eqref{eq:AF_phys_scale}.
Trivially, a narrower mainlobe implies finer range or velocity resolution.

To characterize the sidelobe behavior of each waveform, the \acf{PSLR} and \acf{ISLR} metrics are evaluated, quantifying respectively the peak and total sidelobe energy relative to the mainlobe.  
While these measures are conventionally computed over the entire two-dimensional \ac{AF} surface, in our article, we evaluate them specifically along the \textit{zero-Doppler} and \textit{zero-delay} cuts, only.
This one-dimensional formulation isolates the waveform behavior in the pure delay and Doppler domains, offering a clearer and more interpretable assessment of range- and velocity-related ambiguity characteristics that are intuitively most relevant for \ac{ISAC} analysis.

Given the above, the \ac{PSLR}, commonly measured in dB scale, along a given 1D cut in the $\xi$-domain (where $\xi \in \{\tau, \nu\}$) is defined as
\begin{equation}
\mathrm{PSLR}_{\xi}
= 20\log_{10}
\!\left(
\frac{\max_{\xi \in \text{sidelobes}} |A(\xi)|}
{\max_{\xi} |A(\xi)|}
\right),
\label{eq:PSLR_cut}
\end{equation}
which describes the relative height of the highest sidelobe to the central mainlobe peak, such that lower (more negative) values correspond to stronger sidelobe suppression and lower false-target probability.  

On the other hand, the \ac{ISLR} in dB scale, is defined as the ratio between the total energy in the sidelobes to the energy in the mainlobe, given by
\begin{equation}
\mathrm{ISLR}_{\xi}
= 10\log_{10}
\!\left(
\frac{\sum_{\xi \in \text{sidelobes}} |A(\xi)|^{2}}
{\sum_{\xi \in \text{mainlobe}} |A(\xi)|^{2}}
\right),
\label{eq:ISLR_cut}
\end{equation}
where a smaller \ac{ISLR} indicates greater energy concentration within the mainlobe and reduced energy leakage.  

In the \ac{ISAC} context, these normalized metrics collectively capture the fundamental sensing properties of a waveform: the 3\,dB width determines range or velocity resolution, the \ac{PSLR} measures resistance to false detections, and the \ac{ISLR} reflects the overall energy leakage.
Therefore, in practice, these metrics well describe the waveform's ability and efficiency to accurately estimate target range and velocity in the presence of multiple reflections, which is critical for \ac{6G} scenarios.

\section{Simulation Results and Analysis}
\label{sec:results}

Based on the above waveform formulations and the \ac{AF} analysis framework, we now present and discuss the numerical results comparing the \ac{AF} characteristics of the four candidate waveforms: \ac{OFDM}, \ac{OTFS}, \ac{AFDM}, and \ac{CP-AFDM}. 
Owing to the normalized formulation, these results are fully system-agnostic and can be directly translated to any practical system configuration, enabling fair and consistent comparison across different waveform types.
Furthermore, the smoothing and interpolation techniques yield high-resolution \ac{AF} surfaces and cuts, enabling precise extraction of the key ambiguity metrics, independent of the underlying discrete sample size $N$.\footnote{To promote reproducibility and support future \ac{ISAC} waveform research, the MATLAB simulation scripts for the fractionally-interpolated ambiguity functions are publicly available at our repository: \href{https://github.com/eric-hs-rou/fractional_AF_generator}{{[\textit{online}]}}.}

Two signaling scenarios are considered in the analysis.  
The first corresponds to the general {random-symbol} case, where the transmitted symbols are streams of information, are drawn from a complex modulation alphabet.  
This setting reflects practical \ac{ISAC} operation, in which the waveform must simultaneously convey information and perform sensing. 
Here, the ambiguity function represents the average behavior under typical data transmission, offering a realistic assessment of sensing performance during concurrent communication.  

However, the inherent randomness of the transmitted symbols influences the correlation structure of the waveform and consequently alters the \ac{AF} shape and associated metrics. 
Furthermore, this mode is only valid for sensors (i.e., receivers) that have perfect knowledge of the transmitted symbols in order to perform the matched correlator to obtain the radar responses, which is not always the case in practical \ac{ISAC} scenarios, unless a monostatic configuration is assumed.

Therefore, to assume a more general multi-static case, and also to reveal these underlying core characteristics of the waveform structure independent of data randomness, a second {unimodular} or \textit{sensing-only} scenario is also examined, in which all transmitted symbols are set to unity.  
As an example, such sensing-only operation can be realized in practice by allocating specific time-frequency resources or pilot blocks for dedicated sensing transmissions within an \ac{ISAC} frame.

\subsection{AF Under Random Symbols}

\begin{figure}[H]
\vspace{-2ex}
\centering
\includegraphics[width=1\columnwidth]{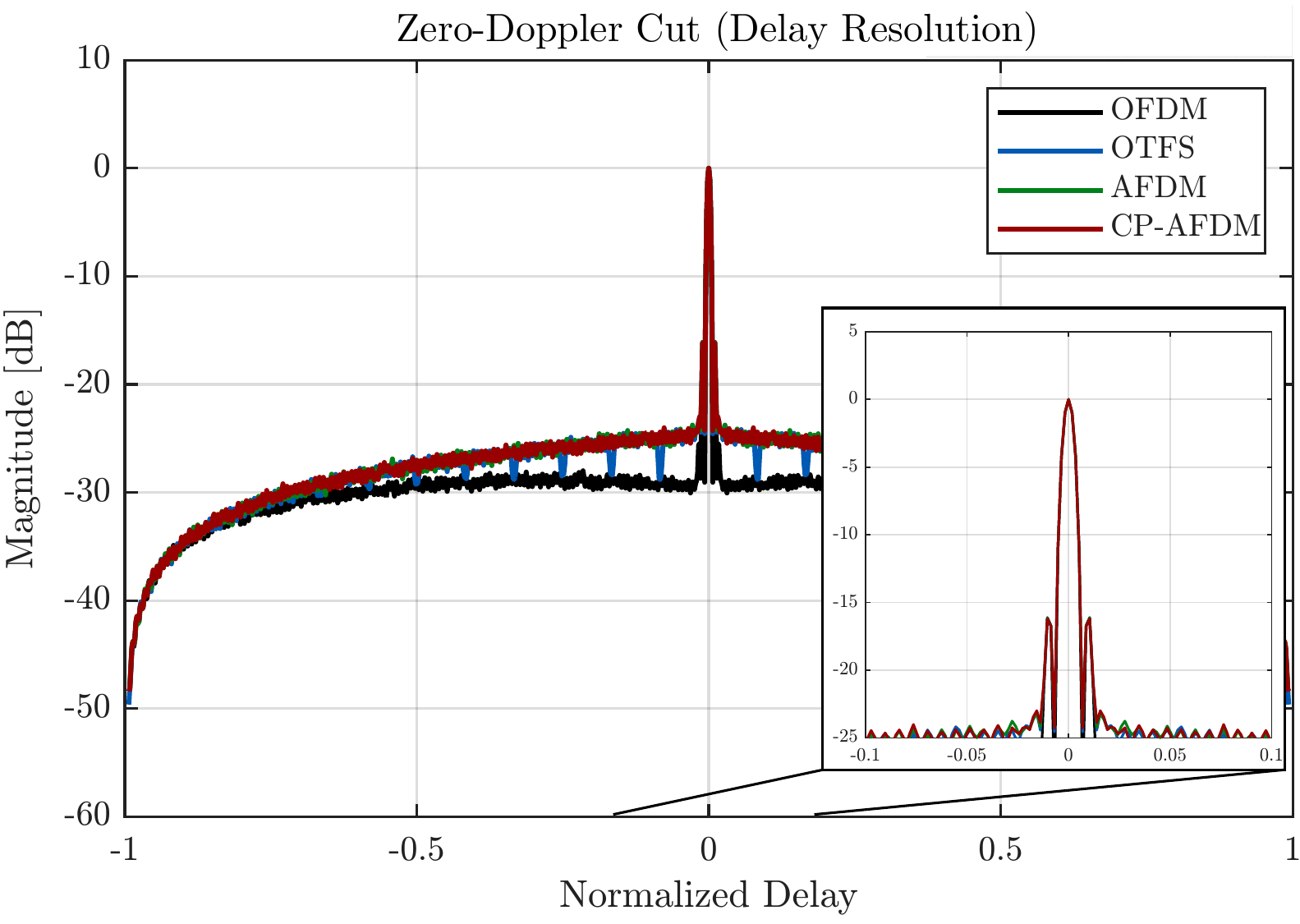}
\vspace{-3.5ex}
\caption{Average zero-Doppler AF cuts for random 16-QAM signaling.}
\label{fig:delay_rand}
\vspace{-3ex}
\end{figure}

\begin{figure}[H]
\vspace{-0.5ex}
\centering
\includegraphics[width=1\columnwidth]{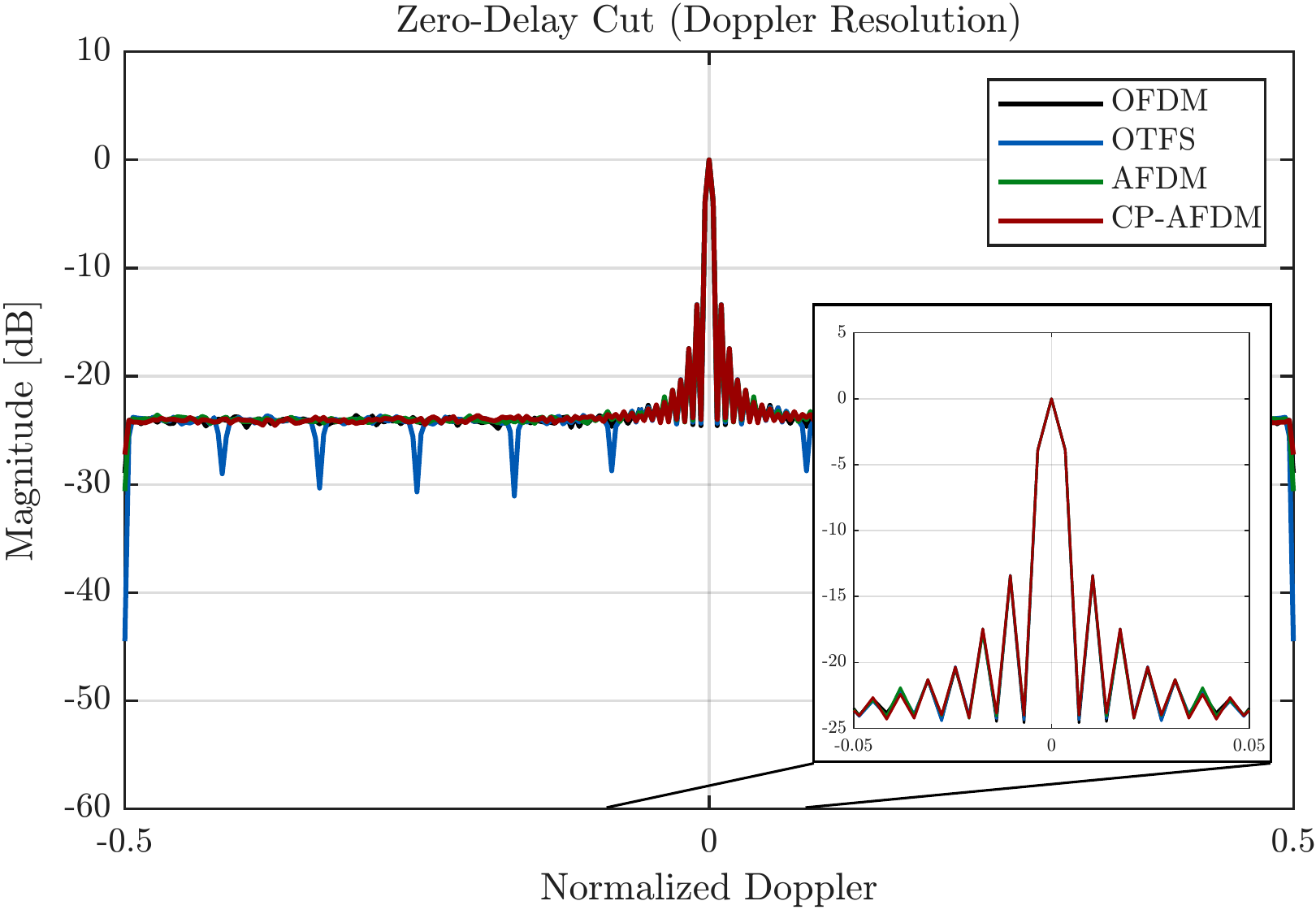}
\vspace{-3.5ex}
\caption{Average zero-delay AF cuts for random 16-QAM signaling.}
\label{fig:dopp_rand}
\end{figure}

Figures~\ref{fig:delay_rand} and~\ref{fig:dopp_rand} illustrate the corresponding \acp{AF} of the four candidate waveforms, with random-symbol averages over $R=1\times10^4$ independent realizations of normalized 16-QAM symbols.
For the \ac{CP-AFDM}, the permutation index $i \in \{1,\ldots,N!\}$ was also randomized for each realization, to reflect the average behavior across all possible permutations, although different permutations can yield slightly different instantaneous \ac{AF} shapes.
Block sizes of $N=144$ symbols for all waveforms ($K \times L = 12 \times 12$ for \ac{OTFS}), with interpolation factors $O_{\tau}=4$ and $O_{\nu}=4$ for delay and Doppler, with $L_h = 4$ have been utilized for the numerical simulations.

As can be seen, the \ac{AF} characteristics are similar across all waveforms, in both the delay and Doppler domains, with the mainlobes being very sharp with low sidelobes, with \ac{OFDM} exhibiting a comparative lower sidelob in the delay domain than other waveforms, as also reported in~\cite{liu2025cp}.
Thanks to the randomness of the symbols, which effectively decorrelates the waveform and suppresses sidelobes, this indicates good delay-Doppler resolution. 
However, as mentioned, this case is only valid for receivers with perfect knowledge of the transmitted symbols (i.e., monostatic \ac{ISAC}), which is not always practical in general \ac{ISAC} scenarios.

\subsection{AF Under Unimodular Symbols}

\begin{table*}[t]
\vspace{1ex}
\centering
\caption{Normalized AF metrics extracted from 1D cuts (Figs. 3 and 4), for the unit symbol case ($N = 144, K = L = 12, O_\tau = 4, O_\nu = 4, L_h = 4$).}
\label{tab:metrics_unimod}
\begin{tabular}{l|ccccccc}
\toprule
Waveform &
$\Delta\tau_{3\mathrm{dB}}$ &
$\Delta\nu_{3\mathrm{dB}}$ &
PSLR$_\tau$ & ISLR$_\tau$ &
PSLR$_\nu$ & ISLR$_\nu$ \\
\midrule
OFDM     & 0.0058 & 1.8021 & -16.0975 & -313.0712 & -14.8007 & -183.6956 \\
OTFS     & 0.0493 & 0.0740 & -43.8373 & ~\,-13.0694 & -51.4935 & ~~~-6.8425 \\
AFDM     & 0.0058 & 0.0056 & ~\,-6.0206 & ~\,-13.4633 & ~\,-0.7274 & ~~~-9.4944 \\
CP-AFDM  & 0.0059 & 0.0058 & -16.0286 & ~\,-14.8963 & ~~\,0.1036 & ~~\,~2.6158 \\
\bottomrule
\end{tabular}
\vspace{-0.5ex}
\end{table*}

\begin{figure}[H]
\vspace{-1ex}
\centering
\includegraphics[width=1\columnwidth]{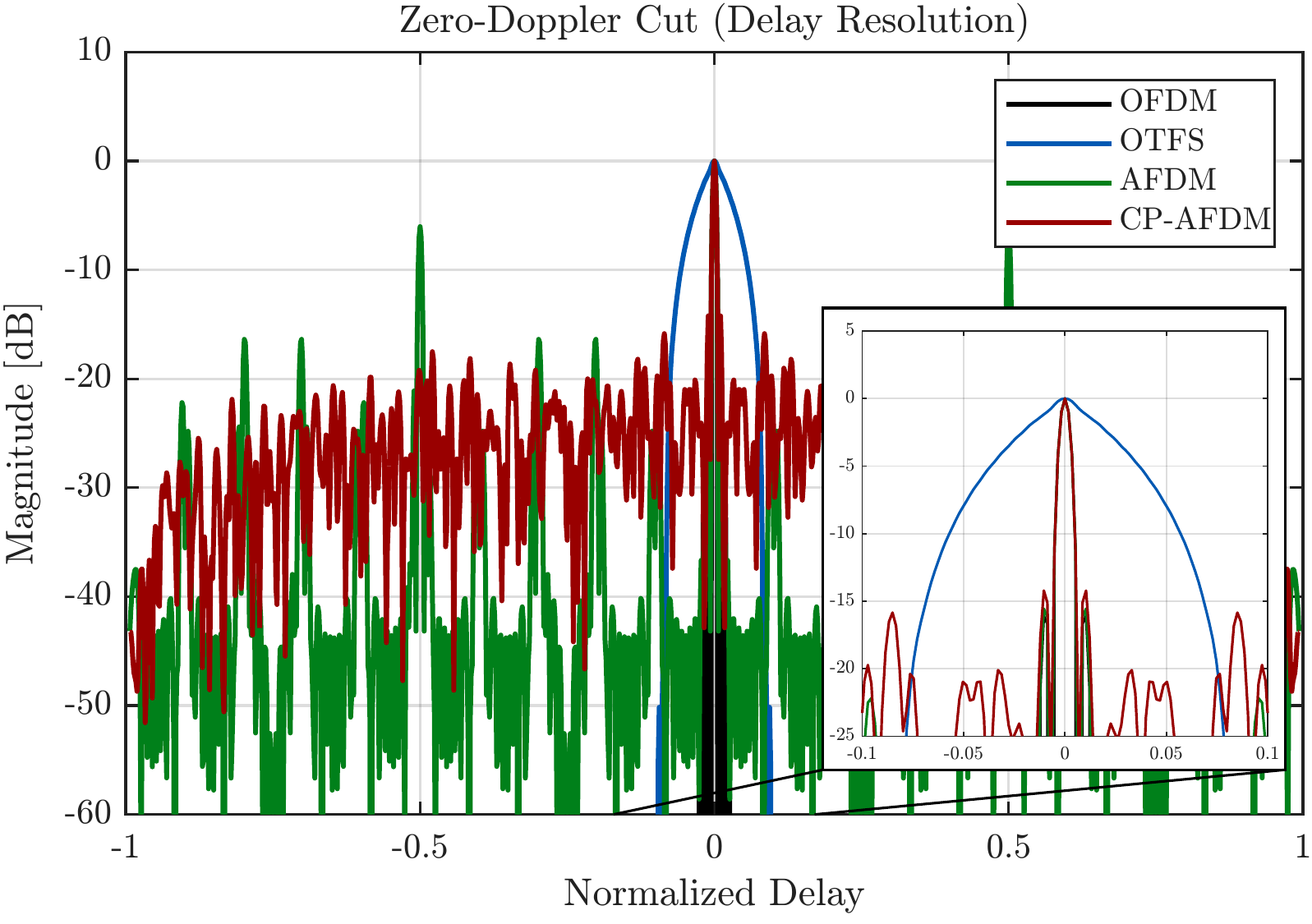}
\vspace{-3.5ex}
\caption{Zero-Doppler AF cuts for unit symbol signaling.}
\label{fig:delay_unimod}
\vspace{-4ex}
\end{figure}

\begin{figure}[H]
\centering
\includegraphics[width=1\columnwidth]{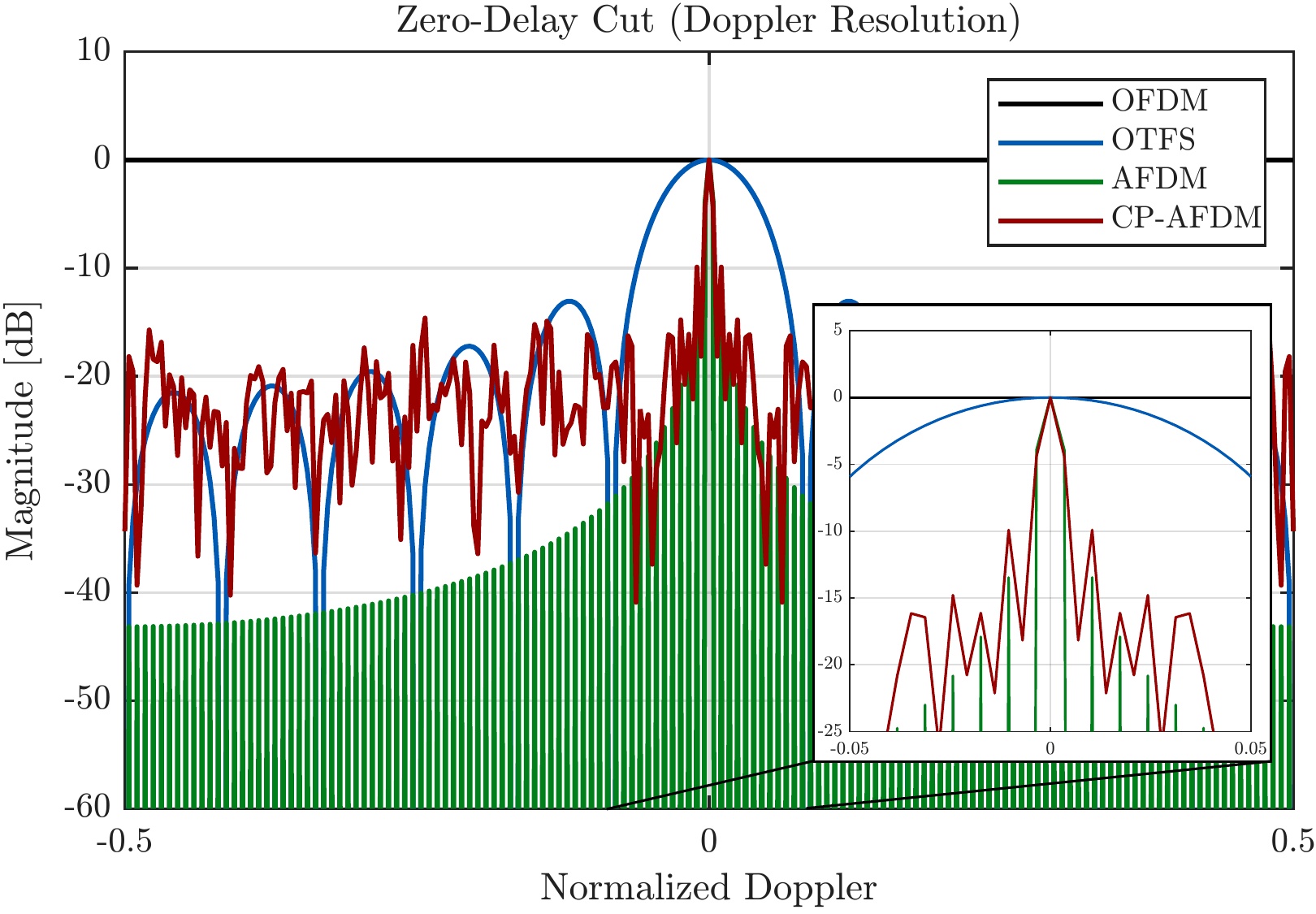}
\vspace{-3.5ex}
\caption{Zero-delay AF cuts for unit symbol signaling.}
\label{fig:dopp_unimod}
\vspace{-0.5ex}
\end{figure}

Figures~\ref{fig:delay_unimod} and~\ref{fig:dopp_unimod} illustrate the normalized zero-Doppler and zero-delay \ac{AF} cuts for the case where all transmitted symbols are set to unity. 
This configuration isolates the intrinsic ambiguity properties of each waveform, independent of symbol randomness, and is relevant for radar-like or multi-static \ac{ISAC} scenarios in which deterministic signals are employed.

In this setting, distinct differences emerge across the four waveforms. 
From the zero-Doppler cuts in Fig.~\ref{fig:delay_unimod}, \ac{OFDM}, \ac{AFDM}, and \ac{CP-AFDM} exhibit sharply concentrated mainlobes with varying sidelobe characteristics, whereas \ac{OTFS} presents a moderately broader mainlobe. 
\ac{OFDM} achieves the lowest sidelobe levels, consistent with its well-known delay resolution properties, while \ac{AFDM} exhibits higher sidelobe peaks around $\pm0.5$ normalized delay, which are effectively mitigated in \ac{CP-AFDM} variant.

Conversely, in the zero-delay cuts of Fig.~\ref{fig:dopp_unimod}, the limited Doppler resolution of \ac{OFDM} becomes apparent, with a nearly flat mainlobe. 
In contrast, both \ac{AFDM} and \ac{CP-AFDM} achieve distinct and narrow Doppler mainlobes, while \ac{OTFS} exhibits a wider but still well-defined response. 
\ac{AFDM} demonstrates the cleanest Doppler-domain behavior with a monotonically decaying sidelobe envelope, followed by \ac{CP-AFDM} and \ac{OTFS}, which display comparatively stronger sidelobes.

It is important to note that the aforementioned behaviors generally hold across all valid parameterizations of the respective waveforms, of course with \ac{OFDM} lacking any adjustable parameter, but \ac{OTFS} on the delay-Doppler lattice dimensions $(K, L)$, \ac{AFDM} on the chirp parameters $(c_1, c_2)$, and \ac{CP-AFDM} on the chirp parameters $(c_1, c_2)$ and additionally on the permutation index $i$. 
This highlights the robustness of the observed ambiguity characteristics and highlights the flexibility of these advanced waveforms in adapting to different configurations while preserving their fundamental delay-Doppler properties.

Finally, to quantitatively support these observations and provide a consistent benchmark of each waveform's intrinsic sensing characteristics, Table~\ref{tab:metrics_unimod} summarizes the extracted normalized \ac{AF} metrics for the unimodular case, including the 3~dB mainlobe widths, \ac{PSLR}, and \ac{ISLR}. 
All quantities are expressed in normalized delay-Doppler units, enabling direct translation to physical parameters such as time delay and Doppler frequency. 
Accordingly, the presented results remain fully system-agnostic and can be readily mapped to any practical configuration of bandwidth, sampling rate, or symbol duration.
It should be noted that although \ac{OFDM} appears to exhibit excellent \ac{PSLR} and \ac{ISLR} in the Doppler domain as well, this effect arises from its excessively wide mainlobe, which makes the sidelobes negligible in relative magnitude. 
Therefore, this is not directly meaningful, as the waveform inherently suffers from extremely poor Doppler resolution.

\section{Conclusion}
\label{sec:conclusion}

We presented a unified, system-agnostic framework to compare the ambiguity functions of \ac{OFDM}, \ac{OTFS}, \ac{AFDM}, and \ac{CP-AFDM} under both random-symbol and unimodular signaling. 
By evaluating the discrete-time AF, applying windowed-sinc delay interpolation and dense Doppler sampling for higher definition, and reporting normalized cut-based metrics (3\,dB width, \ac{PSLR}, \ac{ISLR}), our results enable fair, portable comparisons across systems. 
The methodology provides a concise reference for ISAC waveform selection, highlighting and comparatively illustrating the inherent delay-Doppler characteristics of each candidate waveform.

\end{document}